\documentclass[copyright,creativecommons,sharealike]{eptcs}

\providecommand{\event}{PLACES 2009} 
\providecommand{\volume}{??}
\providecommand{\anno}{??}
\providecommand{\eid}{00}
\usepackage{breakurl}

\usepackage[utf8]{inputenc}
\usepackage{graphicx}
\usepackage{url}
\usepackage{comment}
\usepackage{cite}
\usepackage{color}
\usepackage{capt-of}
\usepackage{tabularx}
\usepackage{listings}

\lstdefinestyle{L}{basicstyle=\ttfamily\scriptsize}

\title{Virtual Machine Support for Many-Core Architectures:\\
Decoupling Abstract from Concrete Concurrency Models}

\author{Stefan Marr$^1$\footnote{Funded by a doctoral scholarship of the 
Institute for the Promotion of Innovation through Science and Technology in Flanders (IWT-Vlaanderen), Belgium.}~, Michael Haupt$^2$, Stijn Timbermont$^1$\footnotemark[1]~, Bram Adams$^3$\\
Theo D'Hondt$^1$, Pascal Costanza$^1$, Wolfgang De Meuter$^1$\\[8pt]
$^1$Programming Technology Lab\\
Vrije Universiteit Brussel, Belgium \\[4pt]
$^2$Hasso Plattner Institute\\
University of Potsdam, Germany\\[4pt]
$^3$Software Analysis and Intelligence Lab\\
Queen's University, Canada}

\def\titlerunning{Virtual Machine Support for Many-Core Architectures}
\def\authorrunning{Marr, Haupt, Timbermont, Adams, D'Hondt, Costanza, and De Meuter}
\def\authorcopyright{
\begin{tabular}[t]{@{}l@{}}Marr, Haupt, Timbermont, Adams,\\ D'Hondt,
  Costanza, and De Meuter\end{tabular}}

\begin{document}
\maketitle

%
\begin{abstract}
The upcoming many-core architectures require software developers to exploit
concurrency to utilize available computational power. Today's
high-level language virtual machines (VMs),
which are a cornerstone of software development, do not
provide sufficient abstraction for concurrency concepts.
We analyze concrete and abstract concurrency models and identify the
challenges they impose for VMs. To provide sufficient concurrency support in VMs,
we propose to integrate concurrency operations into VM instruction
sets.

Since there will always be VMs optimized for special purposes,
our goal is to develop a methodology to design instruction sets with
concurrency support. Therefore, we also propose a list of tradeoffs that
have to be investigated to advise the design of such instruction sets.

As a first experiment, we implemented one instruction set extension
for shared memory and one for non-shared memory concurrency.
From our experimental results, we derived a list of
requirements for a full-grown experimental environment for further
research.
\end{abstract}

\section{Motivation}
With the arrival of many-core architectures, the variance of
processors increases by another order of magnitude. This variance
increases also the need for high-level language virtual machines (VMs) to abstract
from variations introduced by differences among many-core architectures\cite{CellBEA,
Larrabee, GodsonT, Tile64}.
We are concerned with processors having multiple cores, using non-uniform memory access
architectures, and explicit mechanisms for inter-core communication.

For software developers, VMs have to provide abstractions from concrete hardware details
like number of cores or memory access characteristics. In the following subsection,
we categorize three groups of hardware architectures,
which need to be supported by VMs, as \emph{concrete concurrency models}. 
In contrast to those concrete concurrency models, we refer to the concurrency models 
defined by languages or libraries and used by application developers as
\emph{abstract concurrency models}.
Our claim is that the currently available incarnations of abstract concurrency models 
in the form of languages and libraries are not sufficient and need to be complemented
by inherent support for multiple concurrency models by VMs.

To motivate our proposal, we analyze the challenges for VMs with regard to concrete as well as
abstract concurrency models in the remainder of this section.

The remainder of this paper discusses our idea of an instruction set for concurrency
and the research that has to be conducted to develop a methodology which allows
to tailor such an instruction set for the needs of a specific VM and its
application domain.
We give a brief overview of our
initial experiments and present the conclusions for a full-grown experimental
environment. We also discuss the related work which contributs approaches and solutions
to VMs for many-core architectures.

\subsection{Challenges for VMs on Modern Processor Architectures}
\label{sec:ChallendesModernProcessors}

Since processor vendors reached an upper bound on the possible clock speed
to gain more performance, the design of modern processor architectures
diverges from their predecessors in central design elements with each new generation, 
trying to achieve better performance by introducing support for explicit concurrency.

This trend has much different consequences compared to the gradual architectural
changes over the last decade.
Instead of increasing the complexity of the memory hierarchy to hide latency
and bandwidth issues, introducing out-of-order execution of instructions, or
simply raising clock rates, changes are made which are not transparent to software anymore
and require special support. As detailed in the remainder of this section, the \emph{memory access
characteristics} change, the explicit concurrency increases the need for \emph{cache-conscious}
programming, and some architecture introduce explicit inter-core communication which all
needs to be support by VMs.

As already mentioned before, we refer to the concurrency models provided at the
hardware level as \emph{concrete concurrency models}. We identified three models and the challenges they imply for the implementation of VMs.

\subsubsection{Single-core Processor}
The most fundamental \emph{concrete concurrency models} 
is a single-core system accessing memory not shared with
another processor. In such a system, the only notion of concurrency is provided
by the operating system (OS) offering some form of preemptive thread scheduling.

Modern single-core architectures usually use mechanisms
like out-of-order execution of instructions,
vector instructions, or pipeline steps which can lead to parallel execution of small
code portions. However for VM implementations, these forms of parallelism do not impose additional
complexity. It is not necessary to introduce a concurrent garbage 
collector, but a just-in-time (JIT) compiler could still benefit from these mechanisms.

However, for optimal performance, these architectures put another burden
on programmers. Deep cache hierarchies have to be treated carefully for
optimal performance, i.\,e., programmers have to be \emph{cache-conscious}.
Thus, they are responsible for reorganizing data layouts to avoid phenomena like
cache thrashing and support the prefetching heuristics.
JIT compilers could actively use characteristics
like cache line sizes, prefetching heuristics, and branch prediction of the
various hardware architectures for optimization\cite{InstructionScheduling, 
ArtOfMpP}, and interpreters could be adapted, e.\,g., to
assist hardware branch prediction\cite{SuperInstructions}.

With respect to concurrency provided by the OS, a VM has to define
a memory model\cite{1229469} and a task model.
The memory model specifies, amongst others, when a write to a shared variable by one thread can
be seen by reads done by another thread.
These guarantees interact in various ways with JIT compiler optimizations, like
storing temporary values in registers, and OS thread scheduling, since the guarantees
need to be enforced before a thread can be rescheduled. The best performance is 
usually achieved if guarantees are less strong and provide opportunities for
reordering to hide memory latency.

The task model makes concurrency available
to language developers and should allow to schedule tasks with respect to the used data,
to use caches efficiently if tasks, e.\,g., in the form of threads, operate
on shared data.

\subsubsection{Multi-core Processor}
The second concrete concurrency model is a shared memory approach for multi-core
or hardware multi-threaded systems.
To allow a clear distinction to many-core processors (see below), we will concentrate on
systems with an architecture for uniform memory access (UMA)\footnote{Often UMA systems 
are regarded as symmetric multiprocessing (SMP) systems, however, for this
discussion, the memory architecture is the main point of interest and the actually 
utilizations of the cores is subordinated.},
i.\,e., multiple cores or threads connected to a single main memory system
and a cache hierarchy which provides cache coherency.

These architectures have grown from single-core processors and usually share all
important characteristics like deep cache hierarchies and out-of-order execution.
The main difference is the additionally provided hardware concurrency
and cache coherence.

The guarantees given by the memory model are even more important in this case.
Here it is not only arbitrary interleaving but parallel execution which has to
be taken into account.
Overly strict guarantees will require that writes are followed by memory barriers to
ensure that neither instruction-reordering nor the cache hierarchies are hiding
changes at any given time. This will of course hurt performance since both mechanisms
could be practically disabled.

By introducing cache coherency, the appropriate utilization of the available
hardware mechanisms becomes more complex. One example is given by Herlihy and 
Shavit\cite{ArtOfMpP}. They discuss different lock implementations with the basic 
insight that a synchronizing operation like \emph{compare-and-swap} provided by
the processor might hurt performance if used inappropriately. Combined with a simple
read operation which checks whether the value has changed utilizing caching, 
performance can be improved, since relying on cache coherence has less overhead
than an operation which might need to synchronize different cores explicitly
and causes memory operations which cannot be cached.
This insight is not only important for the implementation of synchronization
primitives provided by VMs, but also for the implementation of JIT compilers to
generate efficient code.

Similar to single-core systems, task scheduling should respect data 
dependencies. 
For multi-core systems, scheduling should also be aware of
the cache architecture, i.\,e., how cores share caches and how caches are connected
to a hierarchy, to avoid cache thrashing or rather exploit caching efficiently.

\subsubsection{Many-core Processors}
In contrast to multi-core processors, many-core processors 
cannot rely on a UMA architecture anymore since the known mechanisms do not scale\cite{1467875}.
Instead, these processors rely on non-uniform memory access 
(NUMA) architectures, i.\,e., the cost to access a specific memory location
can be different for all cores.
Furthermore, some architectures will provide explicit communication
facilities between cores and thus will not rely solely on shared memory
for direct communication.
Others will try to avoid this additional complexity. 
However, many-core architectures which provide shared memory and
coherent caches will exhibit performance behavior which will vary with respect to
data locality.

We will discuss three candidates from this category briefly.

\paragraph{Cell BE}
The Cell BE\cite{CellBEA} is already in wide use for media systems as well
as for scientific computing.
One of the major characteristics of the Cell BE is its heterogeneous approach
to combine a central processing element with multiple \emph{synergistic processing
elements} (SPE) to offload computational intensive tasks. 
The SPEs are very simple and are not part of a cache hierarchy, do not feature
out-of-order execution, or even branch prediction. Each one has a local storage
but cannot access main memory directly. Instead, a SPE has to request blocks of
memory to be copied into its local store before it can use the data.

The interconnection of these cores is realized by a ring bus architecture.
Here the physical locality is important to achieve optimal performance.
The ring bus is build from four rings, where two rings can transfer data clockwise
and the other two can transfer data counter clockwise.
A more detailed overview of this architecture is given by Krolak\footnote{\url{http://www.ibm.com/developerworks/power/library/pa-fpfeib/}, Version: 29 Nov 2005}.

\paragraph{TILE64}
The processors produced by Tilera, e.\,g., the TILE64\cite{Tile64} are 
somehow similar to the SPEs with regard to their simplistic design.
However, the TILE64 is a homogenous system with only one type of cores.
Each of the 64 cores has a small cache and is interconnected with neighboring cores
(tiles) via a mesh network with five independent special purpose networks.
Thus, to access memory, a core uses the \emph{memory dynamic network} which transports
the request to the according memory controller and returns the data.
Furthermore, an \emph{inter-cache network} allows to access the local caches of other
cores. Additional inter-core communication networks allow various direct communication
schemes between cores.

The challenges to implement VMs on top of such a system have been documented by
Ungar and Adams\cite{ManyCoreObjectHeap}. The crucial obstacles they encountered
where very small local caches, inefficient communication due to shared memory
(as opposed to explicit core-to-core communication), and
required replication of immutable objects to be cached locally since the processors
cache coherency protocol allows caching of a page only on its home core.
From these observations, we conclude that
adequate strategies will be required to implement object heaps enforced by very small caches, as well
as an appropriate way to harness the available bandwidth for inter-core communication
to reach the theoretical performance maximum. 

\paragraph{Larrabee}
Intel's Larrabee\cite{Larrabee} represents another possible homogenous design.
Similar to the other two designs, the cores itself are much simpler than, for instance,
the latest designs used in desktop computers.
They use an in-order architecture extended by wide vector processing units
since it is primarily designed as a graphics processor.

However, in contrast to the other designs, Intel has decided to go with a cache coherent
system to hide some of the complexity. Each core has its own local subset of
the L2 cache and accesses main memory via the coherent L2 cache using a ring network.
At the moment, it seems that they will not expose this ring network explicitly and
communication is only done via shared memory.
Nonetheless, the performance characteristics will differ drastically from standard
multi-core system especially for systems with more than 16 cores where multiple
short linked rings will be used.

\subsection{Challenges for Abstract Concurrency Models}
\label{sec:MotivationAbstractModels}
Today's \emph{abstract concurrency models} are commonly regarded as not ideal
and a lot research is conducted to improve this situation with different approaches.
In short, shared memory with locking is too complicated \cite{ProblemWithThreads}
and software transactional memory (STM) \cite{STM} as well as Actors \cite{ActorsFormalism,
Actors} are promising but not widely adopted. 

Thus, we expect that ongoing efforts in building languages,
to handle the inherent concurrency of many-core systems,
will likely lead to domain-specific languages and will require support by the underlying
VMs.
In this regard, VMs like the \emph{Java Virtual Machine} (JVM) and the
\emph{Common Language Infrastructure} (CLI) are 
becoming more important as common execution platforms for multiple languages,
since not only the implementation of JIT compilers and efficient garbage
collectors is a tremendous effort, but
the ability to reuse the existing infrastructure surrounding a VM is an 
economical concern.

Realistically, there will not be one single model for expressing concurrency. 
Thus, we argue that a VM has to provide support for a wide range of concurrency models 
at its core. Very likely, developers will have to deal with several models; 
e.\,g., in relation with legacy code requiring proper support. 
Furthermore, support for a wide range of models eases the work of language
designers to implement new ideas or domain-specific solutions. VM developers can also 
benefit from richer concurrency semantics, as it would enable efficient 
implementations of different
abstract concurrency models on top of the concrete models.

To illustrate our argument, we will discuss the example of the actor model \cite{ActorsFormalism}.

The JVM and CLI are both widely used and host all kinds of different programming 
models. Functional as well as imperative languages and in the recent past they started
to provide support for dynamic languages, too.
However, if it comes to concurrency, both support only a shared-memory model with 
threads and locks.

The implementation of an actor-based concurrency model, like it
is found for instance in Erlang\cite{Erlang}, on top of these VMs 
has turned out to be a tough problem.
Karmani et al.\cite{1596658} surveyed different language and library implementations
of actor models on top of the JVM. They observed that only few of them actually 
implement a model which preserves properties like isolation so that actors never share any
state in terms of references to a common object graph.
The few ones which do, usually rely on inefficient mechanisms like serializing the 
object graph which is then send as a copy. A VM could provide support for 
much more efficient zero-copying strategies and enforce the desired properties
of the actor model at the same time.

\subsection{Conclusions}
The presented concrete concurrency models represent actual hardware architectures
which differ widely. The important characteristics are their cache hierarchies,
memory access architectures, the provided form of concurrency,
and means for communication between cores.

Theses characteristics influence not only various implementation details all over
the VM but affect the optimal design of memory, task, and communication model for
each of the different concrete concurrency models. 
For example, the challenge for VMs on many-core architectures is not solely the
utilization of available hardware concurrency but also to use the provided memory
and communication facilities appropriately.
Thus, VMs' concurrency abstraction layers must enable efficient implementations on 
top of the different concrete concurrency models.

To achieve that, 
we argue that VMs should provide explicit and comprehensive support for concurrency.
Explicit support for the various different abstract concurrency models
would allow direct mappings from congruent models which will allow an efficient 
utilization of the available facilities and would ease the task to find a suitable
mapping for the remaining, not directly supported concepts.

For instance, the discussed actor model offers opportunities 
for an efficient mapping onto many-core architectures.
Since cache coherence is an issues in these architectures, it would
be possible to use shared-memory only for immutable global state. 
The state of single actors could be stored in distinct parts of the memory,
so that false sharing is avoided and the small local caches can reach
peak efficiency.
In a standard JVM, it would be rather hard to reconstruct the necessary semantics
for such a mapping from the bytecode, but a semantically enriched instruction set
could would allow a JIT compiler to apply such optimizations.


\section{VM Instruction Sets with Concurrency Support}
\begin{figure}[hbtc]
\begin{center}
\includegraphics[width=0.8\textwidth]{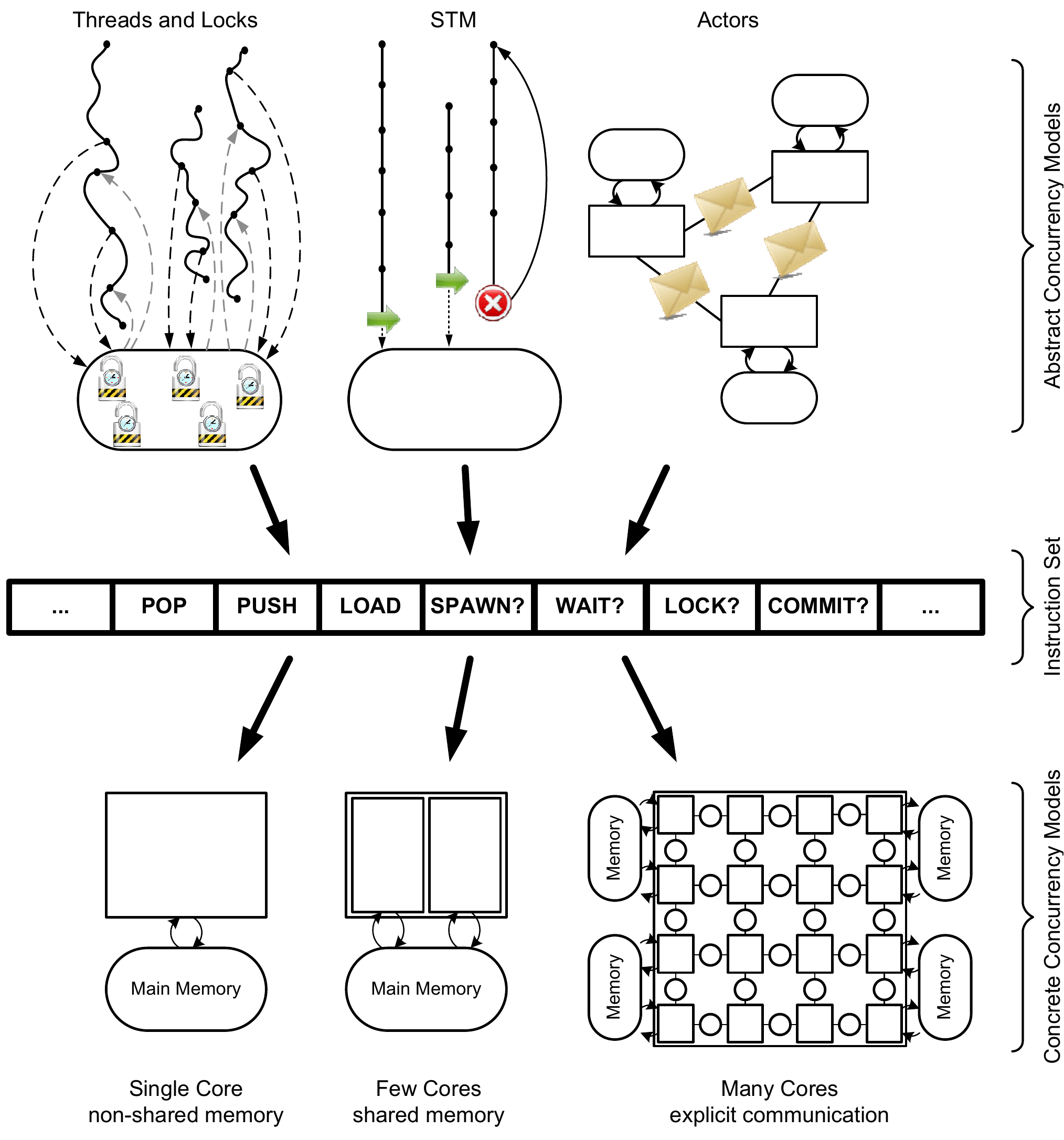}
\caption{A VM instruction set as abstraction layer between abstract and concrete concurrency.}
\label{abstraction-layer}
\end{center}
\end{figure}

Our proposal to achieve a concurrency abstraction layer is to extend the VM instruction
set by concurrency operations. Such an
instruction set will decouple the concurrency models on the different levels of implementation
in such a way that they can be varied independently.
Fig.\,\ref{abstraction-layer} visualizes this idea by showing three different abstract
concurrency models mapped to an instruction set with explicit concurrency support implemented
on top of three different concrete concurrency models.

Expressing concurrency in the instruction set instead of using 
libraries has two major advantages. First,
it will be possible to compile concurrency-related language constructs directly to these 
instructions, avoiding dependencies between languages and libraries on top of the VM.
Second, this choice leads to a larger optimization potential at the VM level, e.\,g., for
JIT compilation, which benefits from the instruction set's precise semantics.

Since there will not be a single instruction set matching all possible requirements, we will work on one 
instruction set representing a very generic
set of requirements, and investigate the design tradeoffs to derive design advice for more concrete 
requirements as well. Thus, we plan to devise
a methodology to develop VM instruction sets with inherent concurrency support, enabling
VM designers to build a concurrency abstraction layer optimized for their
particular requirements.

The methodology will describe how to decouple 
abstract and concrete concurrency. Language designers will be provided with a strategy to 
map abstract concurrency models to instruction sets, and VM implementers
will be enabled to implement instruction sets efficiently on top of concrete
concurrency models. 
The methodology will not only guide such undertakings, but will also give an impression on the effort necessary for their realization.
Below, we discuss our approach in more detail.

\subsection{Approach to Synthesize the Instruction Set}
To devise a broadly applicable methodology, we decided
to adopt a step-wise approach to designing a general instruction set and
discovering the important design tradeoffs. The three currently
most important concurrency models are significantly different in how they represent and realize concurrency: shared-memory with locking, STM, and actors.
For each of these, we will survey different
incarnations in languages or libraries to find each model's 
set of primitives relevant for an instruction set.

Potential candidates for examination are, to name just a few, Java\cite{JavaSpec},
C\#\cite{Ecma334}, Smalltalk\cite{165375}, Cilk/JCilk\cite{Cilk, JCilk}, and frameworks like
Fork/Join for Java\cite{JavaForkJoin}.
Furthermore, constructs like monitors and semaphores are considered as well\cite{506118}.
In the field of STM, we currently consider the work of Shavit and Touitou\cite{STM},
Ziarek et al.\cite{UniformTM}, Saha et al.\cite{1123001}, and Marathe et al. \cite{1066660}.
In the world of actor models the work of Hewitt et al.\cite{ActorsFormalism} and Agha\cite{Actors}
as well as the languages Erlang\cite{Erlang}, Scala\cite{UnifyThreadsEvents},
and Kilim\cite{Kilim} are considered starting points.

\subsection{Ideas to Combine Abstract Concurrency Models}
One of the major research challenges will be to find appropriate combinations of
the different abstract concurrency models. The idea is not to build an 
instruction set which is a simple enumeration of primitives for the different
models, but instead an elaborated combination thereof. Thus, the interaction
between different models has to be completely understood and defined, too.

Our ideas for model combinations are based on the following work.
Volos et al.\,\cite{LocksAndTM} and Blundell et al.\,\cite{UnrestrictedTM} have
described possible solutions for combining locking based code with STM.
A combination of locking based code and actors is described by
Van Cutsem et al.\,\cite{1352693}. STM has many similarities with common transaction processing systems; thus, we will
investigate the application of transaction processing monitors 
\cite{TransactionProcessing} as used in distributed settings to use STM in conjunction with actors. 

\subsection{Tradeoffs to be Investigated}
For the methodology, the discussion of the following design tradeoffs will be an important 
part.

\begin{description}
  \item[Model Combination:] Different solutions to combine concurrency models on the instruction 
         set level will be considered,
	 	 and their benefits and drawbacks investigated. This will reveal
		 critical details like incompatibilities and the possible degree of concurrency.
  \item[Model Mapping:] Strategies to map the concurrency models preserved
         in the instruction set onto concrete concurrency models.
         Here the differences in the memory models, cache hierarchies, and communication
         mechanisms have to be considered.
  \item[Condensed vs. Bloated Instruction Set:] Only few instructions should be added
		 to avoid exceeding the limited number
		 of instructions in a typical bytecode set. However, additional semantics
		 in the instruction set could reduce the complexity of implementing
		 an abstract concurrency model on top of it. It can also be beneficial
		 for an efficient mapping to a concrete concurrency model.
		 Since language and VM implementations should be reasonably manageable,
		 these conflicting interests have to be
		 investigated.
  \item[Bytecode vs. High-level Representation:] Currently,
		 bytecode sets are the most common representation for VM instruction 
		 sets. With respect to communication centric many-core architectures,
		 we will investigate the potential of abstract syntax tree-like high-level
		 representations of interpretable code 
		 in terms of reducing the implementation effort for new instructions
		 and JIT compilers. These investigations will be based on the work of
		 Kistler and Franz\,\cite{TreeJava}.
   \item[Instruction Set vs. Standard Library:] A strategy, to decide which concepts
         are valuable in the instruction set itself, e.\,g.,
         by facilitating JIT compiler optimizations, and which are less common
         or less fundamental and should be provided only in the standard library
         for a given application domain, is necessary, too.
\end{description}

\section{Initial Experiments}

For our first basic experiments we used 
SOM++\footnote{\url{http://hpi.uni-potsdam.de/swa/projects/som/}}, 
a very simple VM, implementing a Smalltalk-like language.
This VM is designed to be used for teaching and to prototype ideas rapidly.

Originally, it has a very small instruction set (16 instructions) and features a
straightforward bytecode interpreter. Its overall design favors simplicity over performance
and utilizes C++ to provide an object-oriented implementation. This results in a VM
implementation which emphasizes conceptual clarity. Thus, experiments usually require
a minimal effort. The downside of this approach is, that SOM++ is considered unoptimized with
regard to performance. Hence, experiments on SOM++ are useful to show the general impact of
different implementation strategies, for instance for garbage collection, but only provide a
rough estimate about performance and interaction effects between subsystems.

In the context of our first experiments, this is not an issue. The goal was to gain an impression
of the general impact of introducing concurrency related instructions into the bytecode
set of a virtual machine. In our experiments, we chose to focus on shared memory and non-shared
memory concurrency in the first place.

The foundation for these experiments is the SOM++ bytecode set. As mentioned before, it consists of 16
instructions. It is purely stack-based and design with simplicity as the main goal in mind.
Thus, the bytecodes are encoded as bytes with the values from 0 to 15. 
Even though it would be possible to encode arguments---e.\,g., indexes for local variables or symbols---within the remaining bits, they are provided as an additional byte each.
Thus, bytecode instruction length varies in the range from 1 to 3. The bytecode set is outlined in Tab.\,\ref{SomBytecodes}.

\begin{table}[htdp]
\begin{center}
\begin{tabularx}{0.95\linewidth}{lX}
\lstinline|DUP| & duplicate top element \\
\lstinline|PUSH_*| & push locals, arguments, fields, blocks, constants, and globals onto stack \\
\lstinline|POP| & remove top element \\
\lstinline|POP_*| & pop top element to locals, arguments, and field variables\\
\lstinline|SEND sig| & send a message identified by \lstinline|sig| to the top element \\
\lstinline|SUPER_SEND| & send a message to the top element, use implementation of the parent class\\
\lstinline|RETURN_LOCAL| & return from the current block of execution to its outer context\\
\lstinline|RETURN_NON_LOCAL| & leave the currently executed method from an inner block\\
\lstinline|HALT| & leave the interpreter loop
\end{tabularx}
\caption{SOM++ bytecode set}
\label{SomBytecodes}
\end{center}
\end{table}

In the following sections, we briefly describe the two experiments, to illustrate potential concurrency
related instructions in VM bytecode sets.

\subsection{Shared Memory Concurrency}
Our very first experiment was to add basic instructions for shared-memory concurrency to
the SOM++ bytecode set. 
We designed the extension similar to the existing instructions. Simplicity was not the foremost
concern here, but we have chosen to add only the five basic instructions outlined in 
Tab.\,\ref{SomThreads}.
They operate on the top element of the execution
stack. For \lstinline|SPAWN|, the top element has to be a block which is then executed in a new thread. As a result, \lstinline|SPAWN| pushes a new thread
object onto the stack. The other four operate on an arbitrary object on the top of 
the stack. The stack itself is not affected.

We relied on an existing implementation of
shared-memory concurrency using the \emph{Pthreads} library. Thus, the largest part of
the work was refactoring the existing implementation from primitives, i.\,e., 
native functions for the Smalltalk thread library to bytecode instructions.
Subsequently, the SOM++ compiler was adapted to emit the new bytecodes on
special messages.

\begin{table}[htdp]
\begin{center}
\begin{tabularx}{0.69\linewidth}{lX}
  \lstinline|SPAWN| & spawn a new thread with the given block on top of the stack \\
  \lstinline|LOCK| & lock the lock of the top element \\
  \lstinline|UNLOCK| & unlock the lock of the top element \\
  \lstinline|WAIT| & wait on a notification on the top element \\
  \lstinline|NOTIFY| & notify all threads waiting on the top element
\end{tabularx}
\caption{Additional instructions for shared memory concurrency}
\label{SomThreads}
\end{center}
\end{table}

In the context of SOM++, the question arose whether it is beneficial to have these instructions in
the instruction set instead of implementing them as primitives. In the course of this project, bytecode instructions are actually the only option, which however brought about considerable overhead in implementing the required extensions in the compiler.

However, SOM++ is not the type of virtual machine we like to target with such 
extensions. Instead,
these kinds of instructions are meant for multi-language virtual machines. Here, the 
purpose of
an instruction set shifts from being a runtime representation of a program to being a 
full-fledged assembly language for all kinds of language implementations.
Thus, a richer instruction set allows to move implementation effort from the
language-level, which has
to be redone for each language, to the platform-level where all language 
implementations can benefit from it without additional effort.

For future experiments we will consider additional shared-memory operations to
increase the flexibility and expressiveness of the instruction set.
At the moment, we think that several low-level operations known from hardware
instruction set architectures could be useful additions to allow language designers
for instance to use lock-free synchronization mechanisms or data structures
at the heard of their languages.

Examples for such operations are atomic updates like \lstinline|XADD|
and \emph{compare-and-swap} (\lstinline|CMPXCHG|) from the IA-32 instruction
set architecture\cite{PentiumSpec}, as well as operations like
\emph{load-and-reserve}/\emph{store-conditional} which are included in the
PowerPC instruction set architecture\cite{PowerPCSpec} in form of \lstinline|lwarx| 
and \lstinline|stwcx|.

\subsection{Non-shared Memory Concurrency}
The second experiment we conducted was inspired by the work of Schippers et\,al.\,\cite{TowardsACMM}
describing an actor-based machine model.
The aim of this experiment was to adapt SOM++ to implement concurrency by actors which do not share
memory, but use explicit message passing for communication.
This kind of machine model is typically found in distributed object systems\,\cite{232711}.

In this model, actors are containers for objects. It is derived from the notion of
\emph{vats} introduced in the E language (and its predecessors)\cite{ELangActors}
where actors are not ``active objects'', but containers for a number of
regular objects.
The contained objects are shielded from undesired concurrent modifications,
since each actor only has a single thread of control.
Messages between actors are exchanged using an incoming message queue per actor.
Objects can reference objects located in another actor by means of remote references.
Usual message sends between objects can be synchronous or asynchronous, independent from
whether the message is sent locally or over a remote reference.

Inside an actor, coroutines are allowed to support a simple means of concurrency.
This is useful since synchronous message sends over remote references
do not block the sending actor, but can yield control to another coroutine until
the return message is received.

To support this machine model, the instruction set had to be adapted as outlined in 
Tab.\,\ref{SomActors}. The basic instructions
stay the same except for \lstinline|SEND|.
For message sends to objects over remote pointers, \lstinline|SEND| was adapted.
It forwards the message sent to the actor owning the object and yields the coroutine to wait
for the result value. The result value is later returned by the \lstinline|RETURN_REMOTE| bytecode.
Usual asynchronous message sends are realized by the \lstinline|SEND_ASYNC| bytecode and
coroutines can explicitly yield control using the \lstinline|YIELD| bytecode. 

\begin{table}[htdp]
\begin{center}
\begin{tabularx}{0.9\linewidth}{lX}
\lstinline|SEND| & sends of remote references yield coroutine and wait for return value \\
\lstinline|RETURN_REMOTE| & sends the return value to the waiting coroutine \\
\lstinline|SEND_ASYNC| & send a message asynchronously to an object, the message
queue of the actor owning the receiving object is used \\
\lstinline|YIELD| & yields control flow, possibly to another coroutine 
\end{tabularx}
\caption{Additional instructions for non-shared memory concurrency}
\label{SomActors}
\end{center}
\end{table}

\subsection{Choosing a Research Platform}
From our experiments, we conclude four requirements
for a full-grown experimental environment fit to
demonstrate the advantages of an instruction set supporting a wide range
of concurrency models:

\begin{itemize}
	\item The VM has to be portable to platforms like TILE64\cite{Tile64}
	      or Cell BE\cite{CellBEA} to be able to evaluate the benefits in
		  mapping from an extended instruction set to different concrete
		  concurrency models.
	\item Implementations of considered abstract concurrency models
          which use a compilation to the VM instruction set as
          implementation strategy should be available.
	\item The VM instruction set should provide space (i.\,e., unused bytecode instructions)
	      for experiments.
	\item The VM should provide an easy to adapt
		  JIT compiler to experiment with optimizations.
\end{itemize}

Based on these requirements, we compiled a list of more than fifty VMs comparing
mainly open source
implementations for various languages like Erlang, JavaScript, Python, and Scheme.
Here we present only a small subset of this comparison to discuss the reasoning
for choosing our research platform.

Tab.\,\ref{VMOverview} lists for each VM the characteristics of interest to choose our
research platform.
The column \emph{language} contains the target language implemented by the VM,
\emph{ACM} reflects the abstract concurrency model.
The availability of threads, STM, and actors implementations are represented by \emph{IS} for
instruction set support, \emph{Lib} for available libraries or language implementations,
or \emph{``-''} if implementations are not available but described in literature.
Furthermore, we consider whether a JIT compiler is available and a port to a many-core system
would be feasible. 
PyPy's thread support is marked with a ``-'', since it relies on a global interpreter lock and
thus does not allow true parallelism. This DisVM was included even so we only have
access to its specification.


\begin{center}
\begin{tabularx}{1.0\linewidth}{llccccccrl}
      &          &           &         &     &        &     &      & Size   & Impl. \\ 
Name  & Language & ACM       & Threads & STM & Actors & JIT & Port & (SLoC) & Lang. \\ \hline
SOM++ & Smalltalk& T/L\footnote{T/L: threads and locks}
                             & IS      &     & IS     &     & x    &     6k & C++ \\
Lua   & Lua      &           & Lib     &     & Lib    &     & x    &    13k & C \\
LuaJIT\footnote{\url{http://luajit.org/}}
      & Lua      &           & Lib     &     & Lib    & x   & x    &    20k & C \\
RVM\footnote{A Squeak VM developed at IBM Research\cite{ManyCoreObjectHeap} for the TILE64}
      & Smalltalk& T/L       & Lib     & -   & -      &     & x    &    28k & C++\\
CacaoVM\footnote{\url{http://www.cacaovm.org/}}
      & Java     & T/L       & Lib     & Lib & Lib    & x   & x    &   121k & C++ \\
Mozart\footnote{\url{http://www.mozart-oz.org}}
      & Oz       & Data-flow &         &     &        &     &      &   159k & C++ \\
Erlang& Erlang   & Actors    &         &     & IS     & x   & x    &   247k & C \\
PyPy  & Python   & T/L       & -       & Lib & Lib    & x   &      &   318k & RPython \\
Maxine\footnote{\url{http://research.sun.com/projects/maxine/}}
      & Java     & T/L       & IS      & Lib & Lib    & x   &      &   361k & Java \\
HotSpot& Java    & T/L       & IS      & Lib & Lib    & x   &      &   540k & C++ \\
JikesRVM& Java   & T/L       & IS      & Lib & Lib    & x   &      &   978k & Java \\
DisVM\footnote{\url{http://doc.cat-v.org/inferno/4th_edition/dis_VM_specification}}
      & Limbo    & CSP\footnote{CSP: Communicating sequential processes\cite{CSP}}
                             &         &     &        &     &      &  spec. &
\end{tabularx}
\end{center}
\captionof{table}{Overview of potential research platforms}
\label{VMOverview}
~\\
Starting with SOM++, we have to conclude from our experience, that its idealized architecture
and its simple implementation allows for fast prototyping of ideas, but on the other hand
might conceal problems associated with our approach especially with regard to performance.

Lua is also small, but has been implemented with a clearer performance objective.
Furthermore, an implementation with a JIT compiler exists
which is small enough to
be ported to a many-core architecture without requiring overly large effort.
Thus, we will consider it as a vehicle to validate our research in the context of
embedded VMs.

The RVM is already tailored to the TILE64 processor.
Since it utilizes the many-core architecture, its special inter-core communication 
facilities, and has a moderate complexity, we will use it for our first experiments,
applying our idea in the setting of many-core systems.

CacaoVM seems to be the smallest and most widely ported open source JVM with a JIT compiler.
Compared to other JVMs in the table, a port of the CacaoVM to a many-core system
should be more feasible, especially since it already has been ported to the 
Cell BE\,\cite{CacaoCell}.

However, it might become necessary to consider VMs like HotSpot, JikesRVM and Maxine when it comes
to the validation of performance properties. At the moment, it is still not clear whether
we will need a JIT compiler with production-level performance to rule out performance
characteristics not introduced by our approach but other modifications done in the development.

Erlang, Mozart, and the DisVM have been included for consideration since they
implement other abstract concurrency models than the usual shared-memory model with 
threads and locks. Interpreted Erlang got already official support for the TILE64
and will allow to conduct partial experiments. However, due to its 
nature of a VM for a functional language and the complexity of its JIT compiler, we will not
chose it as our main research platform.
Mozart implements an abstract concurrency model based on data-flow variables.
Due to its complexity and focus on distributed environments it does not seem to be a feasible
platform for our research.
The DisVM is an interesting design of a VM where the abstract concurrency model is inspired
by CSP. Unfortunately we do not have access to the implementation and thus, an evaluation
as a research platform was not possible.

\section{Related Work}
Support for concurrency in VM instruction sets is currently limited.
The Erlang VM's BEAM instruction set\footnote{\url{http://erlangdotnet.net/2007/09/inside-beam-erlang-virtual-machine.html}}
is a notable exception, providing dedicated support for its efficient light-weight process 
implementation. It includes instructions for asynchronous message sends, reading from the
process' mailbox, waiting and timeouts. It is an example of how one particular model can
be supported at the core of the VM.
Another example is the DisVM. It provides instructions to create channels between non-shared
memory threads as well as to receive and send messages synchronously.
Still, we argue that this concurrency support is not sufficient,
since each VM only provides support for a single abstract concurrency model.
By contrast, today's VMs have to support many different programming models 
to justify the investments in sophisticated and efficient JIT compilers and garbage 
collectors. Thus, they have to provide the basic means for a wide range of concurrency 
models in the same way as they act as execution platforms for different languages.

In the broader field of instruction set design, there are ongoing efforts to extend the
capability of the JVM to act as a platform for different programming languages by introducing
the \texttt{INVOKEDYNAMIC} instruction\footnote{\url{http://jcp.org/en/jsr/detail?id=292}}. More general work on improving instruction sets with semantic extensions\cite{997182, Peymandoust2003} has been done for the hardware level, but the 
concepts for,  e.\,g., compiler adaption can be applied to VMs as well.

For the Cell BE, VM applicability has been evaluated.
Besides porting and designing JVMs for this platform \cite{CellVM, CacaoCell}, 
some optimizations have been considered to utilize available computation power
\cite{1346276, 1366265}.

Distributing a VM over several computational elements
bears additional challenges. Some of them have been addressed for VMs distributed
on cluster setups; e.\,g., class loading, strategies for 
distributed method invocation, data access on the VM level \cite{dJVM}, or thread migration
\cite{Jessica}.

\section{Summary and Future Work}
We proposed to decouple abstract and concrete concurrency models to be able
to cope with the variability of upcoming many-core architectures and their different
memory access architectures.
We argue that this step is necessary to be able to provide support for several kinds of
languages and their abstract concurrency models on top of a VM.
Furthermore, the benefits of a semantically rich concurrency abstraction layer  will allow
more efficient VM implementations on the various different hardware platforms.

The goal of our ongoing research is to design a comprehensive methodology to design VM instruction
sets combining several concurrency models to provide this abstraction.
The methodology will address the various different design tradeoffs.
Our preliminary prototype enabled us to refine our initial requirements for an experimental
environment and provided us with the necessary insights to be able to proceed with our
research on a suitable platform.

The next step of our work is to investigate the design principles for intermediate languages
and the state of the art in concurrency support. Preliminary results on this work have
been presented at the workshop on \emph{Virtual Machines and Intermediate Languages} 2009\cite{VMIL09}.

With the insights of design tradeoffs for the languages, i.\,e., the instruction sets themselves,
we plan to investigate which low-level primitives for shared memory concurrency should be
included. Later, the integration with non-shared memory models in the same language will
be tackled and thus, we will do the step to real multi-model concurrency support for VMs.

\bibliographystyle{eptcs}
\bibliography{references}

\end{document}